# LIMIT CYCLES THAT DO NOT COMPRISE STEADY STATES OF REACTORS

*Marek Berezowski*

*Silesian University of Technology, Faculty of Applied Mathematics*

*44-100 Gliwice, ul. Kaszubska 23, Poland*

*e-mail: marek.berezowski@polsl.pl*

**Abstract**

It is possible that self-induced oscillations appear in reactors, and that their range does not reach the steady state, although such state exists. To prove this, a cascade of tank reactors coupled with mass recycle loop was tested numerically. The above-mentioned phenomenon is characterized by the location of the steady point out of the limit cycle in the phase portrait. This incident may be beneficial to the process, as low steady state does not have to exclude an independent increase of the conversion degree, despite being the only state and not generating oscillations.

## 1. Introduction

The methods of graphic presentation of dynamic phenomena are time series and bifurcation diagrams. In the case of oscillations, the methods also include the so called limit cycles on the phase portrait.

According to available research results, nowadays there is a customary conviction that there should be a steady point inside a limit cycle. If both such solutions are stable, then, depending on the initial conditions, the apparatus can work under the steady state or under the oscillation state (for example: [1] and [2]).

However, this does not always have to be the case. It may happen that the steady point may be set outside the limit cycle. This phenomenon was originally described for a tubular reactor in [3] and [4]. Experimental results of the phenomenon were described by [5]. In this work it was presented on the example of cascade of Continuous Stirred Tank Reactors (cascade of CSTR) (see [6] and [7]) with mass recycle. A large number of tank reactors in the cascade can also applied to approximate a dispersed flow (see [8]). Recycle systems are commonly used in chemical processes, as they enable an increase of the conversion degree and the utilization of the heat from the reaction.

The conducted numerical calculations discussed in the paper indicate that for the assumed model presented below, the setting of the steady point outside the limit cycle takes place already in the case of fifteen reactors. For a smaller number of reactors, this phenomenon rapidly disappears, and then the steady point is surrounded by the limit cycle. The explanation is that the picture visible on the phase plane is only a projection from multi-dimensional space. In the case of a tubular reactor, the projection is from infinitely dimensional space. Thus, we are dealing with a spatial continuum model with infinite number of its own properties. For a cascade consisting of *n* tank reactors, each of which is described by two state variables (concentration and temperature), we are dealing with *2n* dimensional space. It is worth noticing that the discussed phenomenon could also occur for a cascade consisting of two tank reactors.

## 2. The Model

It was assumed in this paper that there are *n* identical pseudo-homogeneous adiabatic chemical tank reactors coupled with the recycle loop (Fig.1).

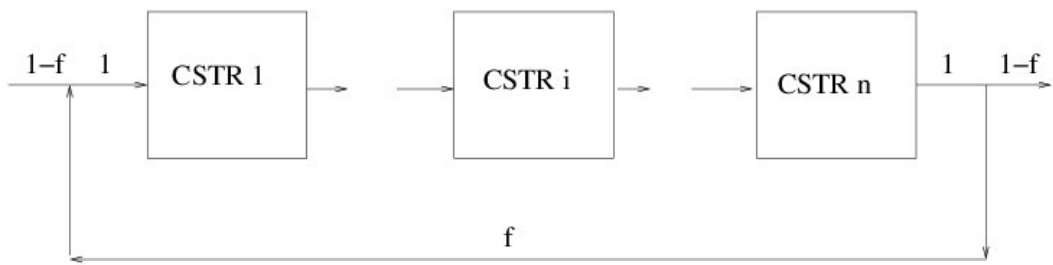

Fig.1. Block diagram of the cascade of tank reactors

The mathematical model for a single reactor is the following:

$$\frac{d\alpha_i}{d\tau} + \alpha_i = \alpha_{i-1} + (1-f)Da(1-\alpha_i)\exp\left(\gamma\frac{\beta\Theta_i}{1+\beta\Theta_i}\right) \quad (1)$$

$$Le\frac{d\Theta_i}{d\tau} + \Theta_i = \Theta_{i-1} + (1-f)Da(1-\alpha_i)\exp\left(\gamma\frac{\beta\Theta_i}{1+\beta\Theta_i}\right) \quad (2)$$

$i = 1,n$, but, due to the recycle, the following boundary conditions have to be fulfilled:

$$\alpha_0 = f\alpha_n; \quad \Theta_0 = f\Theta_n \quad (3)$$

The numerical data assumed for the calculations are: $\gamma = 15$, $\beta = 0.75$, $f = 0.2$, $Le = 10$, $n = 15$. Damköhler number $Da$ was assumed as a bifurcation parameter.

The time series and the limit cycles were derived by means of numerical simulations, whereas, the bifurcation diagram by means of the continuum parametric method, discussed in [7] and [9].

## 3. Results

As the basis for analyzing the above mentioned phenomenon, the bifurcation model illustrated in Fig.2. was assumed.

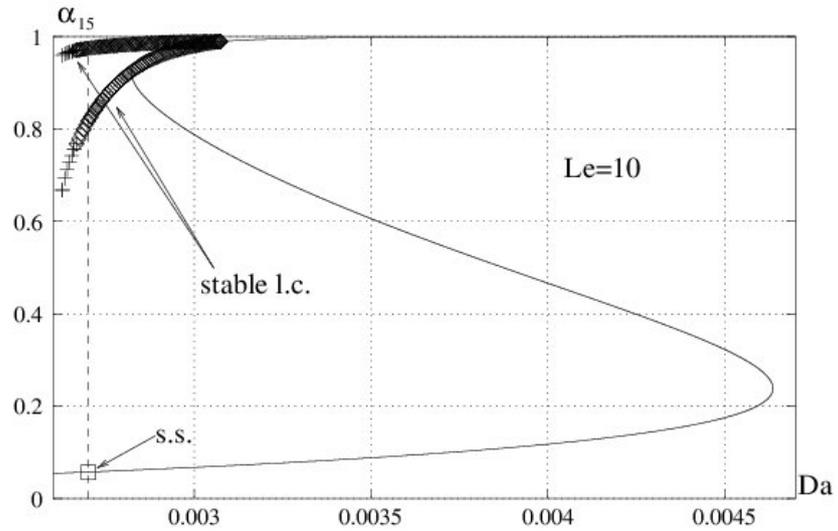

Fig. 2. Bifurcation diagram of the cascade

The variable representing the dynamics of the cascade is the conversion degree of the 15$^{th}$ reactor. The thin line shows the position of the steady states for particular *Da* values. The interval *Da*, in which the so called multiple steady states is also conspicuous. The thick lines mark the amplitudes of stable self-induced periodic oscillations, i.e. stable limit cycles (stable l.c.). It is clear that a fragment of these lines is situated in such a manner that there is no steady state between them. Such phenomenon occurs, for example, for *Da=0.0027*. The only steady point present for this value is situated well below the oscillation branches (s.s.).

In Fig. 3 the time series for stable oscillations are presented (stable l.c.), as well as for unstable oscillations (unstable l.c.).

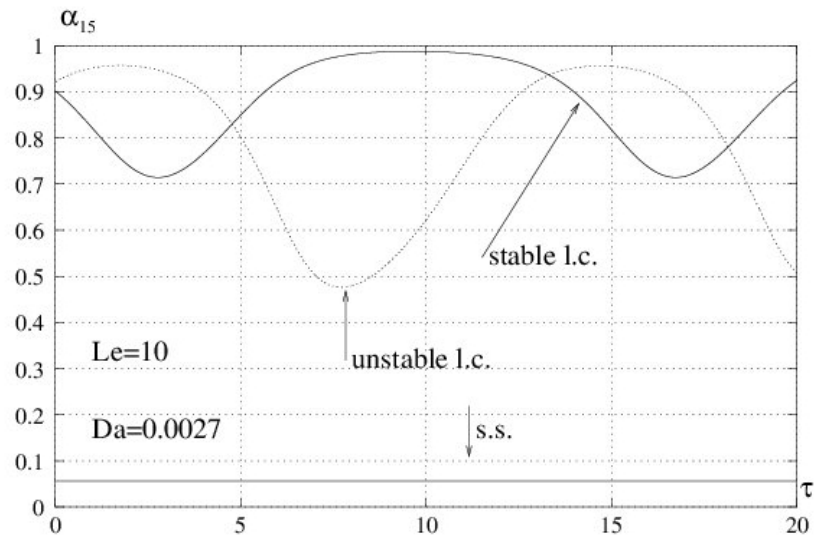

Fig. 3. Time series for the conversion degree

The latter were not presented in the diagram, so as not to obscure the picture. However, it is clearly shown that the oscillation lines do not comprise the steady state, which is marked at the bottom by symbol (s.s). It is characterized by a low conversion degree, which is clearly observable in the phase portrait in Fig. 4, where the above mentioned two limit cycles ( (stable l.c and unstable l.c.) are also marked. The only stable steady point (s.s.) is situated far from these cycles.

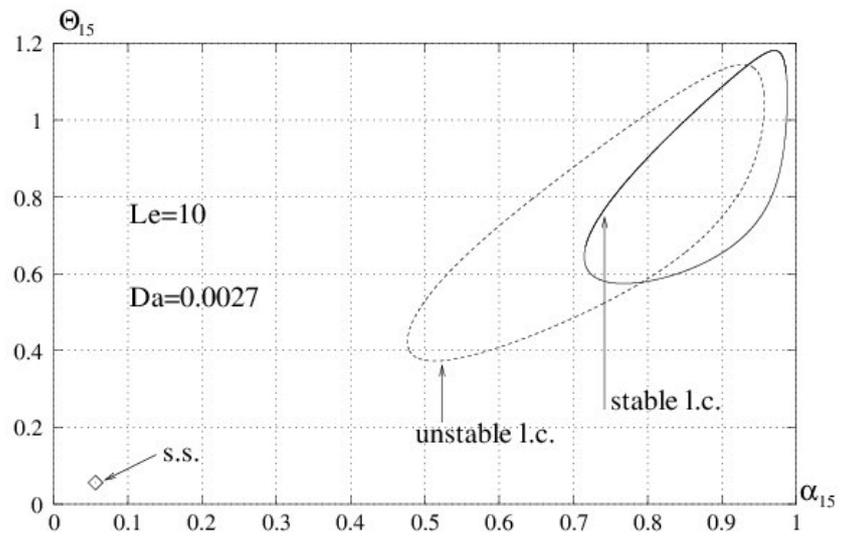

Fig. 4. Phase portrait of the 15<sup>th</sup> reactor

For a more generalized illustration of the discussed phenomenon, in Fig. 5. stable limit cycles (stable l.c.) are shown, as well as stable steady points (s.s.) of the entire cascade, i.e., for all 15 reactors.

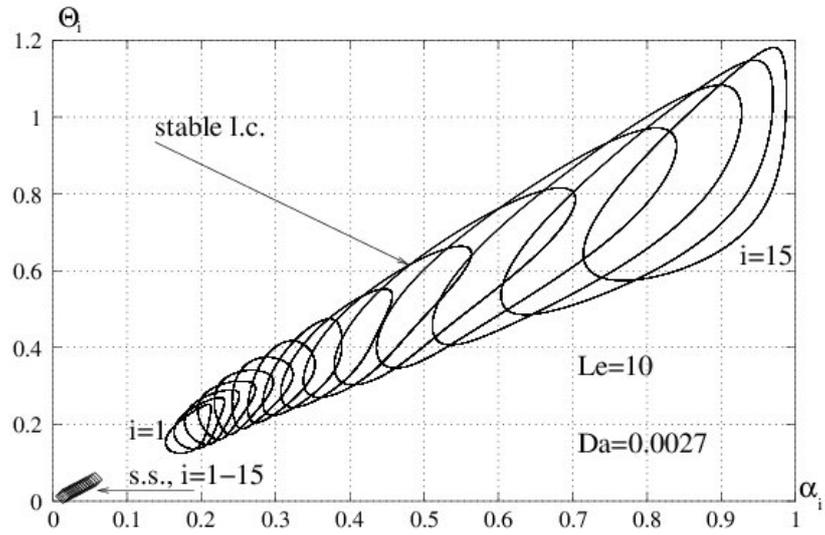

Fig. 5. Phase portrait of the entire cascade

The unstable cycles were not marked, so as not to obscure the picture. It is obvious that not only particular steady points are outside the limit cycles, but, likewise, the entire cascade of the steady points is outside the limit cycles cascade.

To complete the examination of the described phenomenon, a numerical analysis of the influence of Lewis number *Le*. was conducted. As illustrated in Fig.6. starting with a relatively small value of *Le*, the setting of the steady point outside the limit cycle also holds for any values of *Le*.

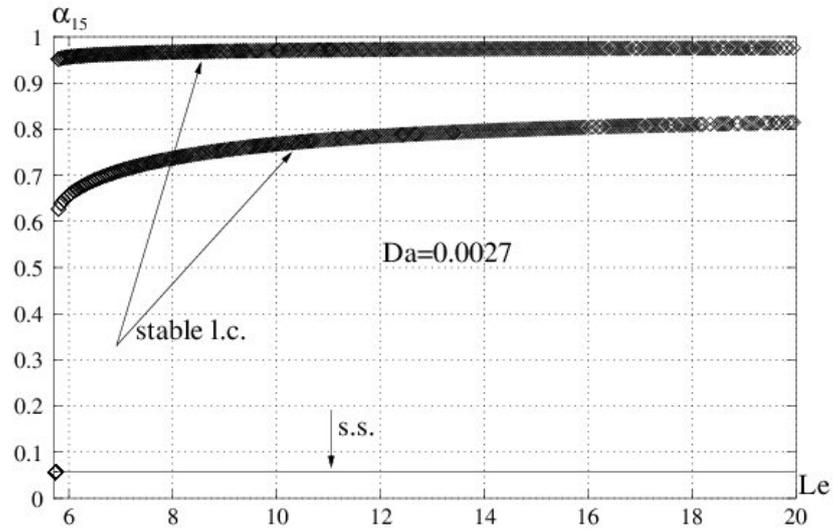

Fig. 6. Influence of *Le* number on the discussed phenomenon

## 4. Summary

The dynamics of the cascade consisting of 15 tank reactors was examined. Moreover, the cascade was coupled with the mass recycle loop. It was confirmed that such system may generate self-induced oscillations. However, the most important accomplishment of this work is the proof that the oscillations do not have to embrace any steady point. This phenomenon is especially well-observed on phase portraits, where the steady point is situated outside the limit cycle. This can be useful at the design stage of a process. A low conversion degree at steady-state response does not exclude a high average degree at oscillation.

**Nomenclature**

| | |
|---|---|
| $c_p$ | heat capacity, *kJ/(kg K)* |
| $C_A$ | concentration of component *A*, $kmol/m^3$ |
| $Da$ | Damköhler number $\left( = \dfrac{V_R(-r_F)}{\dot{F} C_{AF}} \right)$ |
| $E$ | activation energy, *kJ/kmol* |

| | | |
|---|---|---|
| $f$ | recycle ratio, | $\left(=\dfrac{\dot{m}_f}{\dot{m}}\right)$ |
| $\dot{F}$ | volumetric flow rate, $m^3/s$ | |
| $(-\Delta H)$ | heat of reaction, $kJ/kmol$ | |
| $K$ | reaction rate constant, $1/s$ | |
| $Le$ | Lewis number, | $\left(=1+\dfrac{m_w c_{pw}}{m_m c_{pm}}\right)$ |
| $m$ | mass, $kg$ | |
| $\dot{m}$ | mass flow rate, $kg/s$ | |
| $(-r)$ | rate of reaction, $(=KC)$, $kmol/(m^3 s)$ | |
| $R$ | gas constant, $kJ/(kmol\ K)$ | |
| $t$ | time, $s$ | |
| $T$ | temperature, $K$ | |
| $V$ | volume, $m^3$ | |

*Greek letters*

| | | |
|---|---|---|
| $\alpha$ | degree of conversion | $\left(=\dfrac{C_{AF}-C_A}{C_{AF}}\right)$ |
| $\beta$ | dimensionless number related to adiabatic temperature increase | $\left(=\dfrac{(-\Delta H)C_{AF}}{T_F \rho c_p}\right)$ |
| $\gamma$ | dimensionless number related to activation energy | $\left(=\dfrac{E}{RT_F}\right)$ |
| $\Theta$ | dimensionless temperature | $\left(=\dfrac{T-T_F}{\beta T_F}\right)$ |
| $\rho$ | density | $\left(=\dfrac{kg}{m^3}\right)$ |
| $\tau$ | dimensionless time | $\left(=\dfrac{\dot{F}}{V_R}t\right)$ |

*Subscripts*

| | |
|---|---|
| *1* | referes to first CSTR |
| *f* | refers to recycle loop |
| *F* | refers to feed |
| *i* | refers to number of reactor |
| *m* | refers to medium |
| *n* | referes to last CTRS |
| *w* | refers to reactor wall |